\documentclass[sigconf]{acmart}

\usepackage{stfloats}
\AtBeginDocument{%
  }

\setcopyright{acmlicensed}
\copyrightyear{2018}
\acmYear{2018}
\acmDOI{XXXXXXX.XXXXXXX}
\acmConference[SIGCSE TS 2027]{Proceedings of the ACM Technical Symposium on Computer Science Education}{February 17--20, 2027}{Sacramento, CA, USA}
\acmISBN{978-1-4503-XXXX-X/2018/06}

\begin{document}

\title{Understanding Computing Identity Development Through Mentorship and Epistemic Network Analysis}


\author{Behdokht Kiafar}
\email{kiafar@udel.edu}
\orcid{0009-0001-4415-1332}

\affiliation{%
  \institution{University of Delaware}
  \city{Newark}
  \state{DE}
  \country{USA}
}

\author{Roghayeh Leila Barmaki}
\email{rlb@udel.edu}
\orcid{0000-0002-7570-5270}

\affiliation{%
  \institution{University of Delaware}
  \city{Newark}
  \state{DE}
  \country{USA}
}

\renewcommand{\shortauthors}{Anonymous et al.}

\begin{abstract}
Computing identity plays an important role in students’ participation, persistence, and sense of belonging in computing, yet identity development can be difficult to capture through survey measures alone. This study examines how computing identity is expressed in open-ended survey responses from 37 participants in computing-related fields. Using a Quantitative Ethnography approach, we applied Epistemic Network Analysis (ENA) to model co-occurrence patterns among six identity-related constructs: recognition, interest, competence, sense of belonging, self-doubt, and imposter syndrome. We compared the structure of computing identity narratives between participants who reported mentorship support and those who did not. Findings showed that participants with mentorship support had stronger connections among interest, competence, recognition, and sense of belonging, suggesting a more integrated and supportive identity structure. In contrast, participants without mentorship support showed stronger connections involving self-doubt and imposter syndrome, indicating that uncertainty and feelings of not belonging were more closely connected in their narratives. A two-sample t-test comparing ENA scores showed a statistically significant difference between the two groups along the X-axis, with a large effect size (Cohen’s $d = 1.72$). These findings suggest that mentorship is associated with differences in the structure of computing identity and may help individuals connect their interests, abilities, recognition, and belonging within computing.
\end{abstract}




\begin{teaserfigure}
  \centering
  \includegraphics[width=0.9\textwidth]{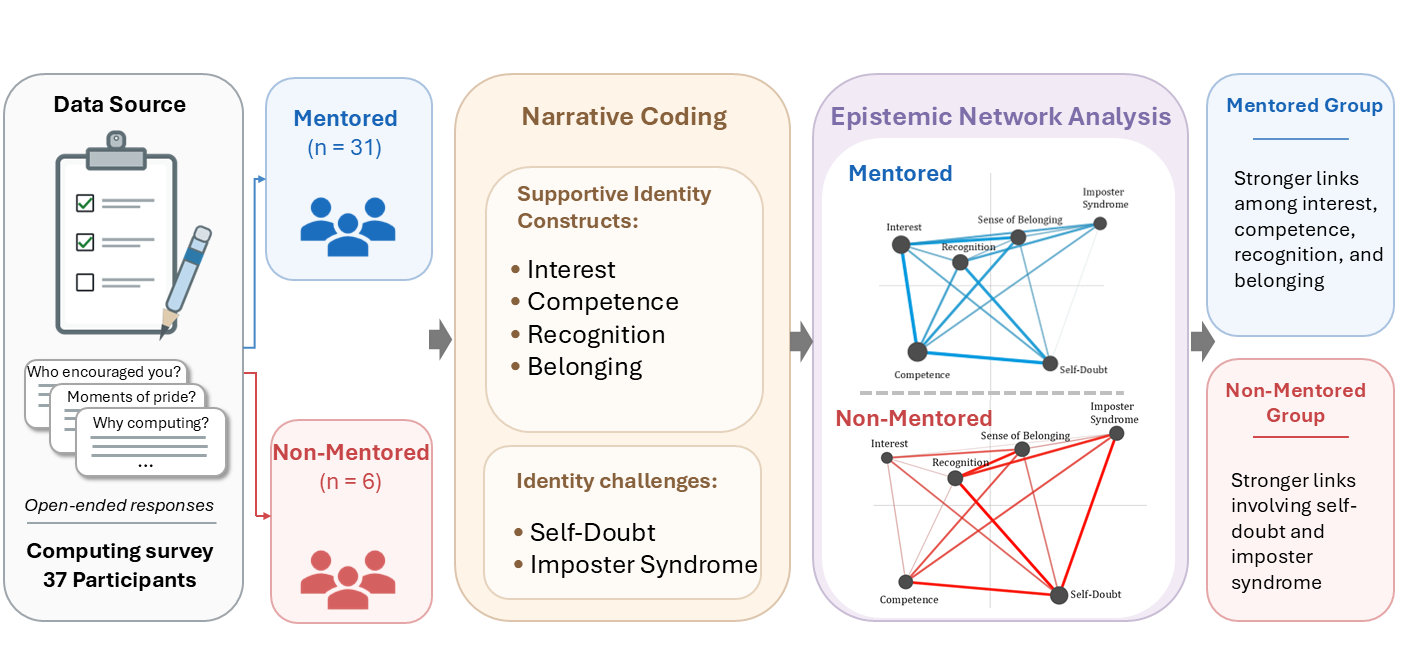}
  \caption{Overview of the study design and key findings. Open-ended survey responses from 37 participants were coded for computing identity constructs and analyzed using Epistemic Network Analysis. Participants with mentorship support showed stronger connections among interest, competence, recognition, and sense of belonging, whereas participants without mentorship support showed stronger connections involving self-doubt and imposter syndrome.}
  \Description{The figure summarizes the study workflow. On the left, the data source is shown as a computing survey with open-ended responses from 37 participants. Participants are divided into a mentored group with 31 participants and a non-mentored group with 6 participants. The responses are then coded for supportive identity constructs, including interest, competence, recognition, and belonging, as well as identity challenges, including self-doubt and imposter syndrome. The coded responses are analyzed using Epistemic Network Analysis, producing separate network visualizations for the mentored and non-mentored groups. The figure shows that the mentored group has stronger connections among interest, competence, recognition, and belonging, while the non-mentored group has stronger connections involving self-doubt and imposter syndrome.}
  \label{fig:teaser}
\end{teaserfigure}


\maketitle

\section{Introduction}

Developing a strong computing identity is an important part of students’ participation and persistence in computing. Students are more likely to see themselves as members of the computing community when they experience interest in computing, perceive themselves as competent, believe that others recognize them as computing people, and feel that they belong in computing spaces \cite{rodriguez2017developing}. These identity-related experiences are especially important because computing pathways often require students to navigate challenging coursework, unfamiliar disciplinary norms, and social environments in which some students may question whether they fit \cite{runa2025sense}. They are also important because students’ pathways into computing may include moments of uncertainty, self-doubt, or imposter feelings that shape how they interpret their ability and belonging. Understanding how students construct and express computing identity is therefore central to efforts to broaden participation and support persistence in computing.

Prior work on disciplinary identity has often conceptualized identity through constructs such as interest, recognition, competence, and performance, while more recent work also emphasizes sense of belonging and identity-related challenges such as self-doubt and imposter syndrome \cite{santhosh2024enriching, taheri2019exploring, rosenstein2020identifying}. These constructs provide a useful foundation for examining students’ relationships to computing. However, identity is not only a set of individual beliefs that can be measured separately; it is also expressed through the ways students connect experiences, motivations, relationships, and self-perceptions when they describe their pathways into and through computing.

Mentorship is one social experience that may shape the development of computing identity. Mentors can provide technical guidance, encouragement, access to opportunities, and recognition that supports students’ confidence and sense of belonging \cite{wofford2023inequitable, boyer2010increasing}.
Yet the relationship between mentorship and computing identity is complex. Mentorship may not simply increase a single identity construct; instead, it may help integrate multiple identity-related constructs, such as interest, competence, recognition, and sense of belonging, while also shaping how students make sense of self-doubt or imposter feelings. This suggests a need for analytic approaches that can examine how identity constructs are connected within students’ own accounts of their experiences.

Many studies of computing identity rely on survey measures to examine relationships among constructs across groups \cite{washington2019ethnic, washington2016computer, taheri2019exploring}. Survey-based approaches are valuable for identifying broad patterns, but they may be less able to capture how participants explain, connect, and make sense of identity-related experiences in their own words. Open-ended survey responses can help address this gap by preserving the relational and contextual nature of participants’ computing experiences. However, these responses can be difficult to compare systematically across participants or groups. This creates a methodological challenge: how can researchers preserve the relational and contextual nature of identity narratives while also producing evidence that can support comparison across meaningful groups?

To address this challenge, this study uses Epistemic Network Analysis (ENA) to examine how computing identity is expressed in open-ended survey responses. ENA is well-suited to this purpose because it models patterns of co-occurrence among coded constructs, allowing researchers to examine not only which constructs appear in participants’ narratives but also how those constructs are connected. Figure~\ref{fig:teaser} provides an overview of the study. In this study, we coded open-ended survey responses from 37 participants in computing fields for identity-related constructs including interest, recognition, competence, sense of belonging, self-doubt, and imposter syndrome. We then used ENA to compare the structure of these responses across participants who reported mentorship experiences and those who did not. To guide this analysis, we asked the following Research Questions (RQs):

\textbf{RQ1.} What patterns of connections among computing identity constructs characterize the narratives of participants with and without mentorship support?

\textbf{RQ2.} How do the computing identity networks of participants with mentorship support differ from those of participants without mentorship support?

Our contributions are as follows:
\begin{itemize}
    \item Collecting open-ended survey data from 37 participants in computing-related fields to examine computing identity, mentorship, challenges, and belonging.
    \item Applying Epistemic Network Analysis to model connections among computing identity constructs.
    \item Comparing participants with and without mentorship support to identify differences in the structure of computing identity responses.
\end{itemize}

By examining computing identity as a network of connected constructs, this work contributes to computing education research in three ways. First, it demonstrates the value of narrative-based and network analytic approaches for studying computing identity. Second, it provides evidence that mentorship is associated with more integrated connections among interest, competence, recognition, and sense of belonging. Third, it offers insight into how mentorship may support computing identity development not only by strengthening individual beliefs, but also by helping participants connect their experiences of support, ability, recognition, and belonging within their broader sense of themselves as computing people.

\section{Background and Related Work}

\subsection{Computing Identity in Computing Education}

Computing identity has become an important construct for understanding how students enter, participate in, and persist within computing pathways. Broadly, identity refers to the ways individuals come to see themselves, and are seen by others, as legitimate participants in a disciplinary community. In computing education, this means more than simply enrolling in computing courses or demonstrating technical skill \cite{kapoor2022categorizing}. Students also develop computing identities through their interests, experiences of competence, opportunities to perform computing practices, and recognition from peers, instructors, mentors, and other members of the computing community \cite{santhosh2024enriching, mahadeo2020developing}.

Prior work on disciplinary identity has often drawn on constructs such as interest, recognition, competence, and performance to explain students’ participation in STEM fields \cite{dou2022constructing, hazari2020context}. Interest reflects students’ curiosity, enjoyment, or personal investment in computing. Competence refers to students’ beliefs about their ability to understand computing concepts or complete computing tasks. Performance captures opportunities to demonstrate computing ability through coursework, projects, internships, research, or other forms of practice. Recognition involves being acknowledged by oneself and others as a “computing person” or as someone who belongs in computing spaces \cite{mahadeo2020developing}. Together, these constructs provide a foundation for understanding how students develop a sense of themselves in relation to computing.

More recent computing education research has extended identity frameworks by emphasizing constructs such as belonging, confidence, persistence, and intersectionality \cite{grosse2023identity}. Sense of belonging is especially important because computing environments can communicate, either explicitly or implicitly, who is expected to participate and succeed. Students who do not see themselves represented in computing, or who experience stereotypes and exclusionary norms, may question whether they belong even when they are academically capable \cite{krause2021relationship}. In addition, identity development may involve challenges such as self-doubt and imposter syndrome, especially when students question their legitimacy, fit within computing spaces, or relative ability to peers \cite{ditton2026relative, feijoo2024navigating}. A student who experiences difficulty in a computing course may view that difficulty as part of learning, or instead as evidence that they are not suited for computing. These interpretations are closely connected to identity development because they influence whether students continue to imagine themselves as future members of the computing field \cite{mogos2021computer}.

Although identity constructs are often studied separately, computing identity is not simply a collection of independent beliefs. Students’ experiences of interest, competence, recognition, belonging, self-doubt, and imposter syndrome are often deeply connected. Attending to these relationships is important because identity development occurs through the ways students connect experiences, relationships, and self-understandings over time \cite{lunn2021educational}. This study therefore treats computing identity as a relational construct: not only what students say about themselves in computing, but also how identity-related ideas are connected within their narratives.

\subsection{Mentorship and Identity Development} 

Mentorship is one important social mechanism through which computing identity may develop. Mentors can provide students with technical guidance, encouragement, academic advice, professional socialization, access to opportunities, and forms of recognition that support students’ participation in computing \cite{boyer2010increasing, wofford2023inequitable}. Such support may be especially important in computing environments where students question their competence, legitimacy, or sense of belonging.

Mentorship may also shape how students interpret challenges. Computing has long been shaped by narrow cultural images of who is perceived as naturally suited for technical work, and these images can affect students’ confidence and participation \cite{mooney2018computer}. Mentors can challenge these messages by affirming students’ abilities, helping them access resources, and making pathways through computing more visible. In this way, mentorship may help students understand difficulty as part of learning rather than as evidence that they do not belong \cite{wofford2023inequitable}.

Prior research suggests that mentorship can support persistence in computing \cite{davis2023equitable}, but the mechanisms through which mentorship shapes identity remain underexplored. In particular, mentorship may not simply increase one identity construct in isolation. Instead, mentorship may help students connect multiple identity-related experiences. For instance, a mentor’s recognition may strengthen a student’s confidence, which may then support perseverance through difficult coursework. Similarly, access to a supportive mentor may increase belonging, which may reinforce interest and commitment to computing. Examining these connections can provide a more nuanced account of how mentorship contributes to computing identity development.


\subsection{Epistemic Network Analysis and Identity}

Epistemic Network Analysis is a quantitative ethnographic method designed to model the structure of connections among coded elements in discourse, interaction, or other forms of qualitative data \cite{shaffer2016tutorial}. ENA is based on the idea that meaning is not only reflected in the presence of individual codes, but also in the patterns of association among codes. By modeling relationships among codes, ENA allows researchers to examine how concepts are connected within participants’ accounts and to compare those patterns across individuals or groups \cite{kiafar2023quantitative, kiafar2024analyzing, kiafar2024enhancing, kiafar2025mena, kiafar2025multimodal, kiafar2025quantitative, shaffer2018epistemic}.

This relational focus has made ENA useful in education research, particularly for studying complex forms of learning, reasoning, collaboration, and professional practice \cite{zhang2022understanding, zhao2024epistemic, zorgHo2023using}. Its value lies in its ability to preserve the relational structure of qualitative data while also producing visual and statistical representations that support comparison. In the context of computing identity, ENA is useful because identity development is relational and multidimensional. By using ENA to compare connections among identity-related constructs across participants with and without mentorship support, we examine mentorship not only as a reported source of support, but also as part of the relational structure through which participants make sense of their computing identity.

\section{Methods and Materials}

\subsection{Data Collection}

\begin{table*}
\centering
\caption{Coding scheme for computing identity analysis, informed by prior theoretical work \cite{mahadeo2020developing, taheri2019exploring, santhosh2024enriching, rosenstein2020identifying}. }
\label{tab:codebook}
\begin{tabular}{p{0.22\linewidth} p{0.55\linewidth}}
\hline
\textbf{Code} & \textbf{Definition} \\
\hline
Recognition & Being acknowledged, encouraged, or seen by others as capable in computing. \\
Interest & Curiosity, enjoyment, or motivation toward computing or technology. \\
Competence & Perceived ability to understand, learn, or perform computing tasks. \\
Sense of Belonging & Feeling included, accepted, or connected within computing spaces. \\
Self-Doubt & Uncertainty or lack of confidence about one's computing ability. \\
Imposter Syndrome & Feeling undeserving, unqualified, or like a fraud despite achievements. \\
\hline
\end{tabular}
\end{table*}

The study protocol was reviewed and approved by the authors’ Institutional Review Board (IRB) prior to data collection. All participants provided consent before completing the study. Recruitment was conducted through email invitations sent to alumni and students in the Computer Information Science department. Participants did not receive compensation. The analytic sample included 37 individuals in computing-related fields, including 31 who reported having mentorship support and 6 who reported no mentorship support.
Participants ranged in age from 20 to 76 years old ($M = 45.4$, $SD = 20.6$). The sample included 13 women and 24 men, with degree levels ranging from some college to doctorate or professional degrees.


Data were collected through an online survey that included demographic questions, short-answer items, and open-ended questions focused on participants’ experiences with computing. These prompts asked participants to describe their pathways into computing, sources of encouragement, challenges encountered, moments of pride or excitement, experiences supporting others, perceptions of mentorship, and views on the inclusiveness of computing environments.

\subsection{Data Formatting and Coding}

\begin{table*} [b]
\centering
\small
\caption{ Examples of de-identified survey response excerpts and binary ENA code assignments.}
\label{tab:coded-examples}
\begin{tabular}{p{0.12\textwidth} p{0.42\textwidth} c c c c c c}
\hline
\textbf{Mentorship Status} & \textbf{Raw Response Excerpt} 
& \textbf{Rec.} & \textbf{Int.} & \textbf{Comp.} & \textbf{Belong.} & \textbf{Self-Doubt} & \textbf{Imp. Syn.} \\
\hline

Mentored &
``My mentor encouraged me to try more challenging projects and told me that I was becoming good at solving programming problems. That made me more interested in continuing in computing.''
& 1 & 1 & 1 & 0 & 0 & 0 \\

\hline

Mentored &
``Being included in the lab meetings and having someone check in on my progress made me feel like I was actually part of the computing community.''
& 1 & 0 & 0 & 1 & 0 & 0 \\

\hline

Non-Mentored &
``I liked programming, but when I got stuck I usually felt like everyone else understood things faster than I did, and I started questioning whether I was really capable.''
& 0 & 1 & 0 & 0 & 1 & 0 \\

\hline
\end{tabular}

\vspace{2mm}
\begin{flushleft}
\small
\textit{Note:} A value of $1$ indicates that the code was present in the response excerpt, and $0$ indicates that it was absent. Rec. = Recognition; Int. = Interest; Comp. = Competence; Belong. = Sense of Belonging; Imp. Syn. = Imposter Syndrome.
\end{flushleft}
\end{table*}

Correct formatting and coding of the data are necessary for applying ENA to model connections among constructs. In ENA, codes are used to categorize meaningful concepts in qualitative data and examine how those concepts co-occur \cite{vandenberg2021prompting}. Therefore, the survey responses were organized into an ENA-ready format, with each row representing one participant’s response to one open-ended survey question.
Each prompt was treated as a stanza, meaning that codes appearing within the same response were considered conceptually connected for the purpose of ENA \cite{shaffer2014formatting}.

We developed a deductive codebook grounded in prior work on computing identity \cite{mahadeo2020developing, taheri2019exploring, santhosh2024enriching, rosenstein2020identifying}. The coding scheme included six constructs: recognition, interest, competence, sense of belonging, self-doubt, and imposter syndrome. Table \ref{tab:codebook} presents the codebook used for this analysis. 
Each response was coded for the presence or absence of each code, with $1$ indicating presence and $0$ indicating absence.
To illustrate this process, Table \ref{tab:coded-examples} shows examples of de-identified response excerpts and their assigned binary codes.
Responses could receive multiple codes when more than one identity-related construct appeared in the same response.

Two raters independently coded the data using the codebook. Inter-rater reliability was calculated using Cohen’s kappa \cite{mchugh2012interrater} and percentage agreement. The initial agreement between the two raters was $89.8$\%, and Cohen’s kappa was $0.816$, indicating strong agreement between the raters. Disagreements were resolved through discussion before preparing the final coded dataset for ENA.
In total, the final coded dataset included $731$ response lines across 37 participants.

\subsection{Applying Epistemic Network Analysis}

In this study, we applied Epistemic Network Analysis to the coded data using the ENA Web Tool (version 1.7.0) \cite{marquart2018epistemic}. Each participant served as the unit of analysis, and participants were grouped by mentorship status: participants who reported having one or more mentors were included in the Mentored group, and participants who reported no mentor support were included in the Non-mentored group.

For each participant, ENA calculated co-occurrences between code pairs within each response and accumulated these connections into a network representation. In the network visualizations, node size represents the relative frequency with which each code appeared, while edge thickness represents the relative strength of co-occurrence between code pairs. The model normalized networks across participants to account for differences in response length before applying singular value decomposition (SVD) for dimensionality reduction \cite{zorgHo2023using}.
SVD produces orthogonal dimensions that maximize the variance explained by the model. The first dimension is represented as $SVD1$, or the $X-axis$, and the second dimension is represented as $SVD2$, or the $Y-axis$. 

The resulting ENA space positioned each participant according to the structure of their code connections. Each participant's projected position in this space is referred to as an ENA score, which summarizes that participant's overall pattern of code connections as coordinates on the SVD dimensions. Fixed node positions allowed direct comparison of group-level networks.
To compare participants with and without mentorship support, we examined group networks, subtraction networks, and group centroids. We then conducted a two-sample t-test assuming unequal variance to test whether the Mentored and Non-Mentored groups differed along the $X$ and $Y$ dimensions of the ENA space.


\section{Results and Discussion}
\subsection{RQ1: Computing identity networks by mentorship status}

\begin{figure*}[t]
    \centering
    \includegraphics[width=\textwidth]{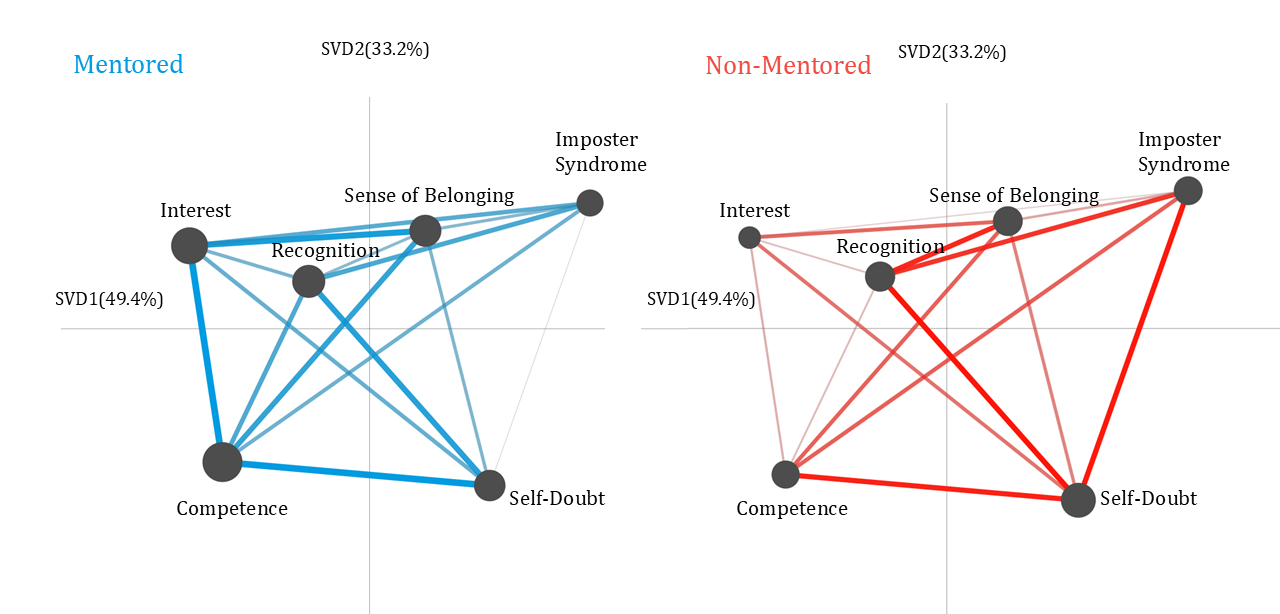}
    \caption{ENA network representations of computing identity constructs by mentorship status. Nodes represent coded identity constructs, with larger nodes indicating more frequent code occurrence, and edges represent the relative strength of co-occurrence between code pairs. Thicker edges indicate stronger connections. The Mentored group shows stronger connections among interest, competence, recognition, and sense of belonging, whereas the Non-Mentored group shows stronger connections involving self-doubt and imposter syndrome.}
    \Description{Two side-by-side ENA network graphs compare participants with mentorship support and participants without mentorship support. Both graphs include six nodes: Interest, Competence, Recognition, Sense of Belonging, Self-Doubt, and Imposter Syndrome. In the Mentored group, the strongest connections appear among Interest, Competence, Recognition, and Sense of Belonging. In the Non-Mentored group, the strongest connections involve Self-Doubt and Imposter Syndrome. Edge thickness represents the relative strength of co-occurrence between constructs.}
    \label{fig:ena-networks}
\end{figure*}

To answer RQ1, we created ENA network models to examine how computing identity constructs were connected in participants’ narratives for the Mentored and Non-Mentored groups. Figure \ref{fig:ena-networks} shows the group-level networks for both groups. The first dimension of the ENA model, $SVD1$, explained $49.4\%$ of the variance, while the second dimension, $SVD2$, explained $33.2\%$ of the variance. Together, these two dimensions captured a substantial portion of the variation in code co-occurrence patterns.

In the Mentored network, the strongest connections appeared among Interest, Competence, Recognition, and Sense of Belonging. Competence was also the largest node in this network, indicating that it appeared most frequently in the Mentored group’s responses. In particular, the strong connection between Interest and Competence suggests that participants with mentorship support often connected their motivation for computing with their perceived ability to learn or perform computing tasks. The visible connections between Interest and Sense of Belonging further suggest that interest in computing was closely tied to feeling included and connected within computing spaces. These patterns indicate that mentorship may be associated with a more integrated structure of supportive identity constructs, in which interest, ability, recognition, and belonging reinforce one another.

The Mentored network also included connections involving Self-Doubt, particularly with Competence and Recognition. This suggests that participants with mentors did not necessarily describe their computing pathways as free of uncertainty or challenge. Rather, self-doubt appeared within a broader network that also included supportive identity constructs. This pattern may indicate that mentorship helps participants contextualize uncertainty as part of learning and development, rather than as evidence that they do not belong in computing.

In contrast, the Non-Mentored network showed stronger connections involving Self-Doubt and Imposter Syndrome. Self-Doubt was the largest node in this network, indicating that uncertainty about computing ability was especially prominent in the Non-Mentored group’s responses. The connection between Self-Doubt and Imposter Syndrome was especially visible, suggesting that participants without mentorship support more often connected uncertainty about their computing ability with feelings of being undeserving, unqualified, or like a fraud. The Non-Mentored network also showed connections between Imposter Syndrome, Recognition, and Self-Doubt. This pattern suggests that, in the absence of mentorship, recognition may be more closely entangled with negative identity-related concerns, including uncertainty about one's ability and feelings of being undeserving or unqualified.

\subsection{RQ2: Differences between Mentored and Non-Mentored identity networks}

To answer RQ2, we compared ENA scores between the Mentored and Non-Mentored groups using a two-sample t-test assuming unequal variance. Along the $X-axis$, the Mentored group ($M = -0.36$, $SD = 0.40$, $N = 31$) differed significantly from the Non-Mentored group ($M = 0.36$, $SD = 0.51$, $N = 6$), $t(6.25) = -3.27$, $p = 0.01$. Cohen’s $d = 1.72$ indicated a large effect size. No significant difference was observed along the $Y-axis$. This result suggests that the primary difference between the groups was captured along the $X-axis$, which separated a more supportive identity structure from a structure more strongly involving self-doubt and imposter syndrome.

\begin{figure}[t]
\centering
\includegraphics[width=\columnwidth]{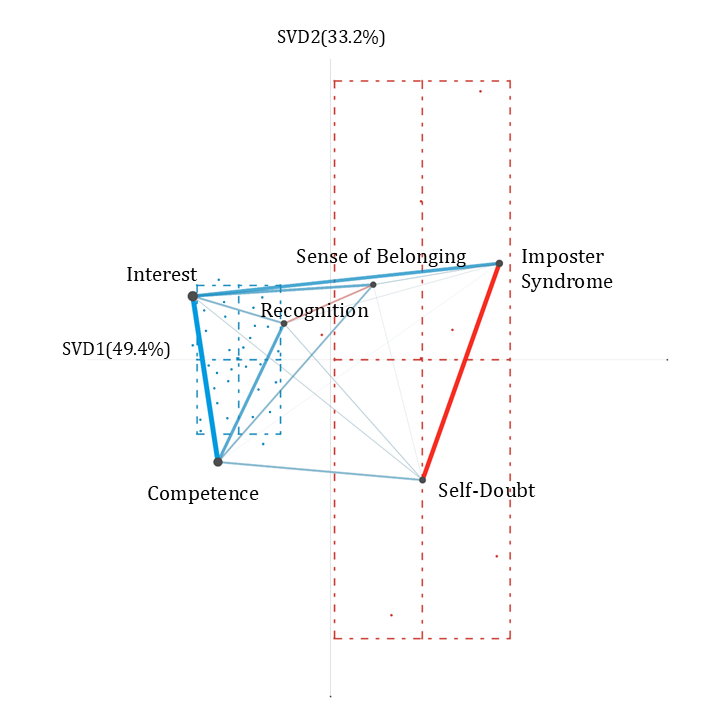}
\caption{
Subtraction network comparing Mentored and Non-Mentored participants. Blue edges show connections stronger in the Mentored group, while red edges show connections stronger in the Non-Mentored group. Mentorship is associated with stronger links among positive identity constructs, whereas lack of mentorship is associated with stronger links involving self-doubt and imposter syndrome.}
\Description{A subtraction network plot with six computing identity constructs: Interest, Competence, Recognition, Sense of Belonging, Self-Doubt, and Imposter Syndrome. Blue edges, mostly on the left side of the network, connect Interest, Competence, Recognition, and Sense of Belonging, showing connections that were stronger for Mentored participants. Red edges, mostly on the right side of the network, connect Self-Doubt, Imposter Syndrome, and Sense of Belonging, showing connections that were stronger for Non-Mentored participants. Dashed blue and red rectangles represent group-level confidence intervals. The main takeaway is that mentorship is associated with stronger positive identity connections, while lack of mentorship is associated with stronger connections involving self-doubt and imposter syndrome.}
\label{fig:subtraction_network}
\end{figure}

Figure~\ref{fig:subtraction_network} presents the subtraction network comparing the Mentored and Non-Mentored groups. This visualization highlights which code connections were relatively stronger for each group by subtracting the connection weights of one group network from the other. Blue edges represent connections that were stronger among participants with mentorship support, while red edges represent connections that were stronger among participants without mentorship support. Each point represents an individual participant’s ENA score. The blue and red squares represent group centroids, and the boxes around these centroids show $95\%$ confidence intervals for each dimension.
The subtraction network provides a more direct view of the connections driving the group difference.

Overall, the ENA results suggest that mentorship is associated with meaningful differences in how computing identity constructs are connected. Participants with mentorship support showed a more integrated network of supportive identity constructs, while participants without mentorship support showed stronger connections among identity challenges. These findings suggest that mentorship may support computing identity development not only by providing encouragement or guidance, but also by shaping how individuals connect their interests, abilities, recognition, and belonging when making sense of their place in computing.

\paragraph{Limitations and Future Work}
This study has several limitations. First, participants came from a voluntary convenience sample, and mentorship groups were naturally occurring rather than randomly assigned. As a result, the findings may be affected by self-selection bias and should not be interpreted as causal. Second, the groups were imbalanced in size, with fewer participants in the Non-Mentored group, which may make the results more sensitive to individual variation. Therefore, these findings should be interpreted as exploratory evidence that mentorship is associated with different patterns of computing identity connections. Future work should examine larger and more balanced samples, consider the type and quality of mentorship, and investigate how computing identity develops over time.

\section{Conclusion}

This study examined how participants describe their computing identity through survey responses and how these identity-related connections differ by mentorship status. Using Epistemic Network Analysis, we modeled computing identity as a network of connected constructs: recognition, interest, competence, sense of belonging, self-doubt, and imposter syndrome. The findings suggest that participants with mentorship support showed stronger connections among interest, competence, recognition, and sense of belonging, indicating a more integrated structure of supportive identity constructs. In contrast, participants without mentorship support showed stronger connections involving self-doubt and imposter syndrome. These patterns suggest that mentorship may shape computing identity not only by providing encouragement or guidance, but also by influencing how individuals connect their abilities, recognition, belonging, and challenges within computing. Overall, this work demonstrates the value of using open-ended survey responses and network analytic methods to study computing identity development.

% ...

\newpage

\bibliographystyle{ACM-Reference-Format}
\bibliography{sigconf}

@article{rodriguez2017developing,
  title={Developing the next generation of diverse computer scientists: The need for enhanced, intersectional computing identity theory},
  author={Rodriguez, Sarah L and Lehman, Kathleen},
  journal={Computer Science Education},
  volume={27},
  number={3-4},
  pages={229--247},
  year={2017},
  publisher={Taylor \& Francis}
}

@article{runa2025sense,
  title={Sense of belonging in undergraduate computing students: A scoping review},
  author={Runa, Shamima Nasrin and McCartan, Andrew and Du, Yuhan and Becker, Brett A and Mooney, Catherine},
  journal={Educational Research Review},
  volume={47},
  pages={100683},
  year={2025},
  publisher={Elsevier}
}

@inproceedings{santhosh2024enriching,
  title={Enriching computing identity frameworks: integrating current constructs and unveiling new dimensions for today’s tech-savvy world—a systematic review},
  author={Santhosh, Malavika E and Siby, Nitha and Sellami, Abdellatif and Bhadra, Jolly and Ahmad, Zubair},
  booktitle={Frontiers in Education},
  volume={9},
  pages={1366906},
  year={2024},
  organization={Frontiers Media SA}
}

@inproceedings{taheri2019exploring,
  title={Exploring computing identity and persistence across multiple groups using structural equation modeling},
  author={Taheri, Mohsen},
  booktitle={American Society for Engineering Education (ASEE) Conference Proceedings},
  year={2019}
}

@article{wofford2023inequitable,
  title={Inequitable interactions: A critical quantitative analysis of mentorship and psychosocial development within computing graduate school pathways},
  author={Wofford, Annie M},
  journal={AERA Open},
  volume={9},
  pages={23328584221143097},
  year={2023},
  publisher={SAGE Publications Sage CA: Los Angeles, CA}
}

@inproceedings{boyer2010increasing,
  title={Increasing technical excellence, leadership and commitment of computing students through identity-based mentoring},
  author={Boyer, Kristy Elizabeth and Thomas, E Nathan and Rorrer, Audrey S and Cooper, Deonte and Vouk, Mladen A},
  booktitle={Proceedings of the 41st ACM technical symposium on Computer science education},
  pages={167--171},
  year={2010}
}

@article{mahadeo2020developing,
  title={Developing a computing identity framework: Understanding computer science and information technology career choice},
  author={Mahadeo, Jonathan and Hazari, Zahra and Potvin, Geoff},
  journal={ACM Transactions on Computing Education (TOCE)},
  volume={20},
  number={1},
  pages={1--14},
  year={2020},
  publisher={ACM New York, NY, USA}
}

@article{dou2022constructing,
  title={Constructing STEM identity: An expanded structural model for STEM identity research},
  author={Dou, Remy and Cian, Heidi},
  journal={Journal of Research in Science Teaching},
  volume={59},
  number={3},
  pages={458--490},
  year={2022},
  publisher={Wiley Online Library}
}

@article{hazari2020context,
  title={The context dependence of physics identity: Examining the role of performance/competence, recognition, interest, and sense of belonging for lower and upper female physics undergraduates},
  author={Hazari, Zahra and Chari, Deepa and Potvin, Geoff and Brewe, Eric},
  journal={Journal of Research in Science Teaching},
  volume={57},
  number={10},
  pages={1583--1607},
  year={2020},
  publisher={Wiley Online Library}
}

@article{grosse2023identity,
  title={Identity in higher computer education research: A systematic literature review},
  author={Gro{\ss}e-B{\"o}lting, Gregor and Gerstenberger, Dietrich and Gildehaus, Lara and M{\"u}hling, Andreas and Schulte, Carsten},
  journal={ACM Transactions on Computing Education},
  volume={23},
  number={3},
  pages={1--35},
  year={2023},
  publisher={ACM New York, NY}
}

@inproceedings{krause2021relationship,
  title={The relationship between sense of belonging and student outcomes in CS1 and beyond},
  author={Krause-Levy, Sophia and Griswold, William G and Porter, Leo and Alvarado, Christine},
  booktitle={Proceedings of the 17th acm conference on international computing education research},
  pages={29--41},
  year={2021}
}

@inproceedings{mogos2021computer,
  title={Computer science identity development in diverse student populations: A qualitative study},
  author={Mogos, Yordanos and Ihorn, Shasta},
  booktitle={SIGCSE},
  year={2021}
}

@article{lunn2021educational,
  title={How do educational experiences predict computing identity?},
  author={Lunn, Stephanie and Ross, Monique and Hazari, Zahra and Weiss, Mark Allen and Georgiopoulos, Michael and Christensen, Kenneth},
  journal={ACM Transactions on Computing Education (TOCE)},
  volume={22},
  number={2},
  pages={1--28},
  year={2021},
  publisher={ACM New York, NY}
}

@inproceedings{mooney2018computer,
  title={Computer science identity and sense of belonging: a case study in Ireland},
  author={Mooney, Catherine and Becker, Brett A and Salmon, Lana and Mangina, Eleni},
  booktitle={Proceedings of the 1st international workshop on gender equality in software engineering},
  pages={1--4},
  year={2018}
}

@article{shaffer2016tutorial,
  title={A tutorial on epistemic network analysis: Analyzing the structure of connections in cognitive, social, and interaction data},
  author={Shaffer, David Williamson and Collier, Wesley and Ruis, Andrew R},
  journal={Journal of learning analytics},
  volume={3},
  number={3},
  pages={9--45},
  year={2016}
}

@incollection{shaffer2018epistemic,
  title={Epistemic network analysis: Understanding learning by using big data for thick description},
  author={Shaffer, David Williamson},
  booktitle={International handbook of the learning sciences},
  pages={520--531},
  year={2018},
  publisher={Routledge}
}

@article{zhang2022understanding,
  title={Understanding student teachers’ collaborative problem solving: Insights from an epistemic network analysis (ENA)},
  author={Zhang, Si and Gao, Qianqian and Sun, Mengyu and Cai, Zhihui and Li, Honghui and Tang, Yanling and Liu, Qingtang},
  journal={Computers \& Education},
  volume={183},
  pages={104485},
  year={2022},
  publisher={Elsevier}
}

@inproceedings{zhao2024epistemic,
  title={Epistemic network analysis for end-users: Closing the loop in the context of multimodal analytics for collaborative team learning},
  author={Zhao, Linxuan and Echeverria, Vanessa and Swiecki, Zachari and Yan, Lixiang and Alfredo, Riordan and Li, Xinyu and Gasevic, Dragan and Martinez-Maldonado, Roberto},
  booktitle={Proceedings of the 14th learning analytics and knowledge conference},
  pages={90--100},
  year={2024}
}

@misc{zorgHo2023using,
  title={Using the reproducible open coding kit \& epistemic network analysis to model qualitative data},
  author={Z{\"o}rg{\H{o}}, Szilvia and Peters, Gjalt-Jorn},
  journal={Health Psychology and Behavioral Medicine},
  volume={11},
  number={1},
  pages={2119144},
  year={2023},
  publisher={Taylor \& Francis}
}

@article{marquart2018epistemic,
  title={Epistemic network analysis (version 1.7. 0)[Software]},
  author={Marquart, CL and Hinojosa, C and Swiecki, Z and Eagan, B and Shaffer, DW},
  journal={Available from app. epistemicnetwork. org},
  year={2018}
}

@article{mchugh2012interrater,
  title={Interrater reliability: the kappa statistic},
  author={McHugh, Mary L},
  journal={Biochemia medica},
  volume={22},
  number={3},
  pages={276--282},
  year={2012},
  publisher={Medicinska naklada}
}

@inproceedings{rosenstein2020identifying,
  title={Identifying the prevalence of the impostor phenomenon among computer science students},
  author={Rosenstein, Adam and Raghu, Aishma and Porter, Leo},
  booktitle={Proceedings of the 51st ACM Technical Symposium on Computer Science Education},
  pages={30--36},
  year={2020}
}

@inproceedings{kapoor2022categorizing,
  title={Categorizing research on identity in undergraduate computing education},
  author={Kapoor, Amanpreet and Gardner-Mccune, Christina},
  booktitle={Proceedings of the 22nd Koli calling international conference on computing education research},
  pages={1--13},
  year={2022}
}

@inproceedings{feijoo2024navigating,
  title={Navigating the impostor phenomenon in computer science education: Insights from two major southeastern institutions in the united states},
  author={Feij{\'o}o-Garc{\'\i}a, Pedro Guillermo and de Siqueira, Alexandre Gomes and Rodr{\'\i}guez-Rey, Tomas Delcl{\'a}ux and Omojokun, Olufisayo},
  booktitle={2024 IEEE Frontiers in Education Conference (FIE)},
  pages={1--6},
  year={2024},
  organization={IEEE}
}

@inproceedings{davis2023equitable,
  title={Equitable student persistence in computing research through distributed career mentorship},
  author={Davis, Sloan and Rorrer, Audrey and Grainger, Cori and Hejazi Moghadam, Sepi},
  booktitle={Proceedings of the 54th ACM Technical Symposium on Computer Science Education V. 1},
  pages={94--100},
  year={2023}
}

@inproceedings{washington2019ethnic,
  title={Ethnic identity as a quantitative measurement in computer science education},
  author={Washington, A Nicki and Romanova, Anna},
  booktitle={2019 Research on Equity and Sustained Participation in Engineering, Computing, and Technology (RESPECT)},
  pages={1--1},
  year={2019},
  organization={IEEE}
}

@inproceedings{washington2016computer,
  title={The computer science attitude and identity survey (CSAIS): A novel tool for measuring the impact of ethnic identity in underrepresented computer science students},
  author={Washington, Alicia Nicki and Grays, Shaefny and Dasmohapatra, Sudipta},
  booktitle={2016 ASEE Annual Conference \& Exposition},
  year={2016}
}

@article{shaffer2014formatting,
  title={Formatting data for epistemic network analysis},
  author={Shaffer, DAVID WILLIAMSON},
  journal={Games and Professionals Imulations (Gaps) Technical Report Series},
  volume={1},
  pages={1--15},
  year={2014}
}

@article{vandenberg2021prompting,
  title={Prompting collaborative and exploratory discourse: An epistemic network analysis study},
  author={Vandenberg, Jessica and Zakaria, Zarifa and Tsan, Jennifer and Iwanski, Anna and Lynch, Collin and Boyer, Kristy Elizabeth and Wiebe, Eric},
  journal={International Journal of Computer-Supported Collaborative Learning},
  volume={16},
  number={3},
  pages={339--366},
  year={2021},
  publisher={Springer}
}

@inproceedings{ditton2026relative,
  title={Relative Self-Efficacy in Computer Science Courses},
  author={Ditton, Joseph and Edwards, John},
  booktitle={Proceedings of the 57th ACM Technical Symposium on Computer Science Education V. 1},
  pages={288--294},
  year={2026}
}

@inproceedings{kiafar2023quantitative,
  title={A quantitative ethnographic examination to improve the quality of training for caregivers for a simulated immersive virtual reality setting},
  author={Kiafar, Behdokht and Daher, Salam and Ahmmed, Asif and Thiamwong, Ladda and Barmaki, Roghayeh Leila},
  booktitle={Fifth International Conference on Quantitative Ethnography: Conference Proceedings},
  volume={8},
  pages={81},
  year={2023}
}

@inproceedings{kiafar2024enhancing,
  title={Enhancing nursing assistant attitudes towards geriatric caregiving through transmodal ordered network analysis},
  author={Kiafar, Behdokht},
  year={2024},
  organization={Sixth International Conference on Quantitative Ethnography: Conference~…}
}

@inproceedings{kiafar2024analyzing,
  title={Analyzing nursing assistant attitudes towards geriatric caregiving using epistemic network analysis},
  author={Kiafar, Behdokht and Daher, Salam and Sharmin, Shayla and Ahmmed, Asif and Thiamwong, Ladda and Barmaki, Roghayeh Leila},
  booktitle={International Conference on Quantitative Ethnography},
  pages={187--201},
  year={2024},
  organization={Springer}
}

@inproceedings{kiafar2025quantitative,
  title={A Quantitative Ethnographic Analysis of Caregiver Competencies and Engagement in Augmented Reality Geriatric Simulation},
  author={Kiafar, Behdokht and Daher, Salam and Ahmmed, Asif and Barmaki, Roghayeh Leila},
  booktitle={International Conference on Quantitative Ethnography},
  pages={321--335},
  year={2025},
  organization={Springer}
}

@inproceedings{kiafar2025multimodal,
  title={Multimodal Analysis of Caregiving Interactions in Simulation-Based Training},
  author={Kiafar, Behdokht},
  booktitle={Proceedings of the 27th International Conference on Multimodal Interaction},
  pages={726--729},
  year={2025}
}

@inproceedings{kiafar2025mena,
  title={MENA: A Multimodal Framework for Analyzing Caregiver Emotions and Competencies in AR Geriatric Simulations},
  author={Kiafar, Behdokht and Ravva, Pavan Uttej and Daher, Salam and Ahmmed Joy, Asif and Leila Barmaki, Roghayeh},
  booktitle={Proceedings of the 27th International Conference on Multimodal Interaction},
  pages={181--190},
  year={2025}
}


\end{document}